\documentclass[12pt,preprintnumbers,fullsize]{article}
\advance\hoffset by -1.2truecm
\advance\voffset by -1.5truecm
\usepackage{graphicx}
\usepackage{epsfig}
\usepackage{dcolumn}
\usepackage{bm}

\def\gsim{\lower0.5ex\hbox{$\:\buildrel >\over\sim\:$}}
\def\lsim{\lower0.5ex\hbox{$\:\buildrel <\over\sim\:$}}

\newcommand{\be}{\begin{equation}}
\newcommand{\ee}{\end{equation}}
\newcommand{\beq}{\begin{eqnarray}}
\newcommand{\eeq}{\end{eqnarray}}

\def\bra{\langle}
\def\ket{\rangle}

\def\b{\beta}

\def\G{\Gamma}
\def\d{\delta}

\def\e{\epsilon}

\def\m{\mu}
\def\n{\nu}
\def\p{\pi}
\def\ph{\phi}
\def\r{\rho}
\def\t{\tau}
\def\s{\sigma}

\begin{document}

\begin{flushright}
{TIFR/TH/04-19}\\
{August 2004}
\end{flushright}
\vspace{2cm}

\begin{center}
{\Large{\bf The Charged Higgs Boson Search at LHC}}
\vspace{2cm}

{\large{\bf D. P. Roy}} \vspace{5mm}


{\bf Department of Theoretical Physics} \\
{\bf Tata Institute of Fundamental Research} \\
{\bf Homi Bhabha Road, Mumbai 400 005, India} \vspace{3mm}

{dproy@theory.tifr.res.in}\par\vspace*{3cm}
\end{center}

\begin{abstract}
{This review starts with a brief introduction to the charged
Higgs boson $(H^\pm)$ in the Minimal Supersymmetric Standard Model
(MSSM). It then discusses the prospects of a relatively light $H^\pm$
boson search via top quark decay and finally a heavy
$H^\pm$ boson search at LHC. The viable channels for $H^\pm$
search are discussed, with particular emphasis on
the $H^\pm \to \t \n$ decay channel.}
\end{abstract}


\section{Introduction}

The minimal supersymmetric extension of the Standard Model (MSSM)
contains two Higgs doublets $\ph^{+,0}_u$ and $\ph^{0,-}_d$, with
opposite hypercharge $Y=\pm 1$, to give masses to the up and down type
quarks and leptons. This also ensures anomaly cancellation between
their fermionic partners. The two doublets of complex scalars
correspond to 8 degrees of freedom, 3 of which are absorbed as
Goldstone bosons to give mass and longitudinal components to the
$W^\pm$ and $Z$ bosons. This leaves 5 physical states: two neutral
scalars $h^0$ and $H^0$, a pseudo-scalar $A^0$, and a pair of charged
Higgs bosons $H^\pm$. While it may be hard to distinguish any one of
these neutral Higgs bosons from that of the Standard Model, the
$H^\pm$ pair carry a distinctive hall-mark of the MSSM. Hence the
charged Higgs boson plays a very important role in the search of the
SUSY Higgs sector.


At the tree-level all the MSSM Higgs masses and couplings are given in
terms of two parameters -- the ratio of the vacuum expectation values,
$\tan \b = \bra \ph^0_u \ket / \bra \ph^0_d \ket$, and any one of the
masses, usually taken to be $M_A$. The physical $H^\pm$ and $A^0$
states correspond to the combinations
\beq
H^\pm &=& \ph^\pm_u \cos \b + \ph^\pm_d \sin \b , \nonumber \\ [2mm]
A^0 &=& \sqrt{2} ({\rm Im} \ph^0_u \cos \b + {\rm Im} \ph^0_d \sin \b) ,
\label{eq1}
\eeq
while their masses are related by
\be
M^2_{H^\pm} = M^2_A + M^2_W ,
\label{eq2}
\ee
with negligible radiative corrections \cite{one}. 

The important couplings of the charged Higgs boson are 
\beq
H^+ \bar{t}b &:& {g \over \sqrt{2}M_W} (m_t \cot \b + m_b \tan
\b), \ H^+ \t \n : {g \over \sqrt{2}M_W} m_\t \tan \b, \nonumber \\[2mm]
H^+ \bar{c}s &:& {g \over \sqrt{2} M_W} (m_c \cot \b + m_s \tan \b), 
 H^+ W^- Z : 0 ,
\label{eq10}
\eeq
with negligible radiative corrections. 

The $H^+ \bar{t}b$ Yukawa coupling of eq.(\ref{eq10}) is ultraviolet
divergent. Assuming it to remain perturbative upto the GUT scale
implies
\be
1 < \tan \b < m_t/m_b (\sim 50) .
\label{eq11}
\ee
However this assumes the absence of any new physics beyond the MSSM
upto the GUT scale -- i.e. the socalled desert scenario. Without this
assumption one gets weaker limits from the perturbative bounds on this
coupling at the electroweak scale, i.e.
\be
0.3 < \tan \b < 200 .
\label{eq12}
\ee
Moreover there is a strong constraint on the $M_A - \tan \b$ parameter
space coming from the LEP-2 bound on the $H_{SM}$ mass, which is also
applicable to $M_h$ at low $\tan \b$, i.e. $M_h > 114$ GeV
\cite{two}. Comparing this with the MSSM prediction 
implies $\tan \b > 2.4$ for any value of $M_A$ \cite{one,two} (see
Fig. 2 below).  However the MSSM prediction for $M_h$ depends
sensitively on the top quark 
mass. The recent increase of this mass from 175 to 178 $\pm$ 4.3 GeV
\cite{three} along with a more exact evaluation of the radiative
correction \cite{four} have resulted in a significant weaking of this
constraint. In fact there is no LEP bound on $\tan \b$ now, which
would be valid for all values of $M_A$. Nonetheless it implies $M_A >$
150 GeV ($M_{H^\pm} >$ 170 GeV) over the low $\tan \b (\leq 2)$
region. But being an indirect bound, it depends strongly on the
underlying model. There is no such bound in the CP violating MSSM due
to $h$-$A$ mixing \cite{five}. Moreover there are singlet extensions
of the MSSM Higgs sector like the socalled NMSSM, which invalidate
these $M_A (M_{H^\pm})$ bounds without disturbing the charged Higgs
boson \cite{six}. Therefore it is prudent
to relax these indirect constraints on $M_{H^\pm}$ and $\tan \b$, and
search for $H^\pm$ over the widest possible parameter space. It should
be noted here that the $H^\pm$ couplings of eq.(\ref{eq10}) continue
to hold over a wide class of models. In fact the fermionic couplings
hold for the general class of Type-II two-Higgs-doublet models, where
one doublet couples to up type and the other to down type quarks and
leptons \cite{one}.

\section{Search for a Light $H^\pm (M_{H^\pm} < m_t)$}

The main production mechanism in this case is top quark pair production 
\be
q\bar q, gg \to t\bar t ,
\label{eq13}
\ee
followed by
\be
t \to b H^+ \ {\rm and/or} \ \bar t \to \bar b H^- . 
\label{eq14}
\ee
The dominant decay channels of $H^\pm$ are 
\be
H^+ \to c \bar s, \t^+ \n \ {\rm and} \ W b \bar b + hc , 
\label{eq15}
\ee
where the 3-body final state comes via the virtual $t \bar b$
channel. All these decay widths are easily calculated from the Yukawa
couplings of eq.(\ref{eq10}). The QCD correction can be simply
implemented in the leading log approximation by substituting the quark
masses appearing in the Yukawa couplings by their running masses at
the $H^\pm$ mass scale \cite{seven}. Its main effect is to reduce the
$b$ and $c$ pole masses of 4.6 and 1.8 GeV respectively \cite{two} to
their running masses $m_b (M_{H^\pm}) \simeq $ 2.8 GeV and $m_c
(M_{H^\pm}) \simeq 1$ GeV. The corresponding reduction in the $t$ pole
mass of 175 GeV is only $\sim$ 5 \%.

The $t \to bH^+$ branching ratio is large at $\tan \b \lsim 1$ and
$\tan \b \gsim m_t/m_b$, which are driven by the $m_t$ and the $m_b$
terms of the $H^+ \bar t b$ coupling of eq. (3) respectively. However
it has a pronounced minimum around $\tan \b = \sqrt{m_t/m_b} \simeq
7.5$, where the SM decay of $t \to bW$ is dominant. The $H^\pm$ is
expected to decay dominantly into the $\t\n$ channel for $\tan \b >
1$, while the $cs$ and the $b\bar b W$ channels dominate in the $\tan
\b \leq 1$ region. This can be easily understood in terms of the
respective couplings of eq.(\ref{eq10}). The $H^+
\to \bar b bW$ three-body decay via virtual $t\bar b$ channel is
larger than the $H^+ \to c\bar s$ decay for $M_{H^\pm} \gsim $ 140
GeV, although the former is a higher order process
\cite{eight,nine}. This is because the $H^+\bar t b$ coupling is
larger than the $H^+\bar c s$ coupling by a factor of $m_t/m_c > 100$
in the low $\tan \b$ region.  The $M_{H^\pm} < 140$ GeV region has
already been excluded at $\tan\beta \leq 1$ by the $t \rightarrow b
H^+ \rightarrow bc \bar s$ search at Tevatron [2].  With a much larger
$t\bar t$ production rate at LHC one can extend the search to the
$M_{H^\pm} > 140$ GeV region via the $H^\pm \rightarrow b\bar b W$
channel at $\tan\beta \leq 1$ [9].  Let us concentrate however on the
$H^\pm \rightarrow \tau\nu$ channel, which dominates the theoretically
favoured region of $\tan\beta > 1$.
 
\subsection{$\t$ Polarization Effect} 

The discovery reach of the $\t$ channel for $H^\pm$ search at Tevatron
and LHC can be significantly enhanced by exploiting the opposite
polarization of $\t$ coming from the $H^\pm \to \t\n (P_\t = +1)$ and
$W^\pm \to \t\n (P_\t = -1)$ decays \cite{seventeen}. Let me briefly
describe this simple but very powerful method. The best channel for
$\t$-detection in terms of efficiency and purity is its 1-prong
hadronic decay channel, which accounts for 50\% of its total decay
width. The main contributors to this channel are 
\beq
\t^\pm &\to& \p^\pm \n_\t (12.5\%), \ \ \t^\pm \to \r^\pm \n_\t \to 
\p^\pm \p^0 \n_\t (26\%), \nonumber \\ [2mm]
\t^\pm &\to& a^\pm_1 \nu_\t \to \p^\pm \p^0 \p^0 \n_\t (7.5\%),
\label{eq17}
\eeq 
where the branching fractions of the $\p$ and $\r$ channels include
the small $K$ and $K^\ast$ contributions respectively \cite{two},
which have identical polarization effects. Together they account for
more than 90\% of the 1-prong hadronic decay of $\t$. The CM angular
distributions of $\t$ decay into $\p$ or a vector meson $v (= \r,
a_1)$ is simply given in terms of its polarization as
\beq
{1\over \G_\p} \ {d\G_\p \over d\cos\theta} &=& {1\over 2} (1 + P_\t
\cos\theta), \nonumber \\ [2mm]
{1\over \G_v} \ {d\G_{vL} \over d\cos\theta} &=& {{1\over 2} m^2_\t
\over m^2_\t + 2 m^2_v} (1 + P_\t \cos\theta) , \nonumber \\ [2mm]
{1\over \G_v} \ {d\G_{vT} \over d\cos\theta} &=& {m^2_v \over m^2_\t +
2 m^2_v} (1 - P_\t \cos\theta) ,
\label{eq18}
\eeq
where $L,T$ denote the longitudinal and transverse polarization states
of the vector meson \cite{seventeen,eighteen}. This angle is related
to the fraction $x$ of the $\t$ lab. momentum carried by the meson,
i.e. the (visible) $\t$-jet momentum, via
\be
\cos\theta = {2x - 1 - m^2_{\p,v}/m^2_\t \over 1 - m^2_{\p,v}/m^2_\t} .
\label{eq19}
\ee
It is clear from (\ref{eq18}) and (\ref{eq19}) that the signal $(P_\t
= + 1)$ has a harder $\t$-jet than the background $(P_\t = -1)$ for
the $\p$ and the $\r_L, a_{1L}$ contributions; but it is the opposite
for $\r_T, a_{1T}$ contributions. Now, it is possible to suppress the
transverse $\r$ and $a_1$ contributions and enhance the hardness of
the signal $\t$-jet relative to the background even without
identifying the individual resonance contributions to this
channel. This is because the transverse $\r$ and $a_1$ decays favour
even sharing of momentum among the decay pions, while the longitudinal
$\r$ and $a_1$ decays favour uneven distributions, where the charged
pion carries either very little or most of the momentum
\cite{seventeen,eighteen}. Fig. 1 shows the decay distributions of
$\r_L, a_{1L}$ and $\r_T, a_{1T}$ in the momentum fraction carried by
the charged pion, i.e.
\be
x' = p_{\p^\pm}/p_{\t {\rm -jet}} .
\label{eq20}
\ee
The distributions are clearly peaked near $x' \simeq 0$ and $x' \simeq
1$ for the longitudinal $\r$ and $a_1$, while they are peaked in the
middle for the transverse ones. Note that the $\t^+ \to \p^\pm \n_\t$
decay would appear as a $\d$ function at $x' = 1$ on this plot. Thus
requiring the $\p^\pm$ to carry $>$ 80\% of the $\t$-jet momentum,
\be
x' > 0.8 ,
\label{eq21}
\ee
retains about half the longitudinal $\r$ along with the pion but very
little of the transverse contributions. This cut suppresses not only
the $W \to \t\n$ background but also the fake $\t$ background from QCD
jets\footnote{Note that the $x'\simeq 0$ peak from $\r_L$ and $a_{1L}$ can not
be used in practice, since $\t$-identification requires a hard
$\p^\pm$, which will not be swept away from the accompanying neutrals
by the magnetic field.}. Consequently the $\t$-channel can be used for
$H^\pm$ search over a wider range of parameters. The resulting $H^\pm$
discovery reach of LHC is shown on the left side of Fig.2
\cite{nineteen}. It goes upto $M_A \simeq 100$ GeV ($M_{H^\pm} \simeq$
130 GeV) around the dip region of $\tan \b \simeq 7.5$ and upto $M_A
\simeq 140$ GeV ($M_{H^\pm} \simeq 160$ GeV) outside this region.

\section{Search for a Heavy $H^\pm (M_{H^\pm} > m_t)$}

The main production process here is the leading order (LO) process
\cite{twenty}
\be
gb \to tH^- + h.c. 
\label{eq22}
\ee
The complete NLO QCD corrections have been recently calculated by two groups \cite{twentyone,twentytwo}, in agreement with one another. Their main results are summarized below:
\begin{enumerate}
\item[{(i)}] 
The effect of NLO corrections can be incorporated by multiplying the
above LO cross-section by a $K$ factor, with practically no change in
its kinematic distributions.
\item[{(ii)}] 
With the usual choice of renormalization and factorization scales,
$\m_R = \m_F = M_{H^\pm} + m_t$, one gets $K \simeq 1.5$ over the
large $M_{H^\pm}$ and $\tan\b$ range of interest.
\item[{(iii)}] 
The overall NLO correction of 50\% comes from two main sources --- (a)
$\sim$ 80\% correction from gluon emission and virtual gluon exchange
contributions to the LO process (\ref{eq22}), and (b) $\sim -$ 30\%
correction from the NLO process
\be
gg \to t H^- b + h.c. ,
\label{eq23}
\ee
after subtracting the overlapping piece from (\ref{eq22}) to avoid
double counting.
\item[{(iv)}] 
As clearly shown in \cite{twentytwo}, the negative correction from (b)
is an artifact of the common choice of factorization and
renormalization scales. With a more appropriate choice of the
factorization scale, $\m_F \simeq (M_{H^\pm} + m_t)/5$, the correction
from (b) practically vanishes while that from (a) reduces to $\sim$
60\%. Note however that the overall $K$ factor is insensitive to this
scale variation.
\item[{(v)}] 
Hence for simplicity one can keep a common scale of 
$\m_{F,R} = M_{H^\pm} + m_t$ along with a $K$ factor of 1.5, with an estimated
uncertainty of 20\%. Note that for the process (\ref{eq22}) the
running quark masses of the $H^+ \bar tb$ coupling (\ref{eq10}) are to
be evaluated at $\m_R$, while the patron densities are evaluated at
$\m_F$.
\end{enumerate}

The dominant decay mode for a heavy $H^\pm$ is into the $tb$
channel. The $H^\pm \to \t\n$ is the largest subdominant channel at
large $\tan \b (\gsim 10)$, while the $H^\pm \to W^\pm h^0$ can be the
largest subdominant channel over a part of the small $\tan \b$ region
\cite{one}. Let us look at the prospects of a heavy $H^\pm$ search at
LHC in each of these channels. The dominant background in each case
comes from the $t\bar t$ production process (\ref{eq13}).

\subsection{Heavy $H^\pm$ Search in the $\t\n$ Channel} 

This constitutes the most important channel for a heavy $H^\pm$ search
at LHC in the large $\tan\b$ region. Over a large part of this region,
$\tan \b \gsim 10$ and $M_{H^\pm} \gsim 300$ GeV, we have
\be
BR (H^\pm \to \t\n) = 20 \pm 5 \% .
\label{eq24}
\ee
The $H^\pm$ signal coming from (\ref{eq22}) and (\ref{eq24}) is
distinguished by very hard $\t$-jet and missing-$p_T (p\!\!\!/_T)$,
\be
p^T_{\t{\rm -jet}} > 100 GeV \ {\rm and} \ p\!\!\!/_T > 100 GeV ,
\label{eq25}
\ee
with hadronic decay of the accompanying top quark $(t \to b q \bar q)$
\cite{twentythree}. The main background comes from the $t\bar t$
production process (\ref{eq13}), followed by $t \to b\t\n$, while the
other $t$ decays hadronically. This has however a much softer $\t$-jet
and can be suppressed significantly with the cut
(\ref{eq25}). Moreover the opposite $\t$ polarizations for the signal
and background can be used to suppress the background further, as
discussed earlier. Figure 3 shows the signal and background
cross-sections against the fractional $\t$-jet momentum carried by the
charged pion (\ref{eq20}). The hard charged pion cut of (\ref{eq21})
suppresses the background by a factor of 5-6 while retaining almost
half the signal cross-section. Moreover the signal $\t$-jet has a
considerably harder $p_T$ and larger azimuthal opening angle with the
$p\!\!\!/_T$ in comparison with the background. Consequently the
signal has a much broader distribution in the transverse mass of the
$\t$-jet with the $p\!\!\!/_T$, extending upto $M_{H^\pm}$, while the
background goes only upto $M_W$. Figure 4 shows these distributions
both with and without the hard charged pion cut (\ref{eq21}). One can
effectively separate the $H^\pm$ signal from the background and
estimate the $H^\pm$ mass from this distribution. The LHC discovery
reach of this channel is shown in Fig. 2, which clearly shows it to be
the best channel for a heavy $H^\pm$ search at large $\tan\b$. It
should be added here that the transition region between $M_{H^\pm} >
m_t$ and $< m_t$ has been recently analysed in \cite{twentyfour} by
combining the production process of (\ref{eq22}) with
(\ref{eq13},\ref{eq14}). As a result it has been possible to bridge the gap
between the two discovery contours of Fig. 2 via the $\t\n$ channel.

\subsection{Heavy $H^\pm$ Search in the $tb$ Channel} 

Let us discuss this first for 3 and then 4 b-tags. In the first case
the signal comes from (\ref{eq22}), followed by
\be
H^\pm \to t \bar b, \ \bar t b .
\label{eq26}
\ee
The background comes from the NLO QCD processes 
\be
gg \to t \bar t b \bar b, \ gb \to t \bar t b + h.c., \ gg \to t\bar t g ,
\label{eq27}
\ee
where the gluon jet in the last case can be mistagged as $b$ (with
a typical probability of $\sim 1\%$). One requires leptonic decay of
one of the $t\bar t$ pair and hadronic decay of the other with a $p_T
> 30$ GeV cut on all the jets \cite{twentyfive}. For this cut the
$b$-tagging efficiency at LHC is expected to be $\sim$ 50\%. After
reconstruction of both the top masses, the remaining (3rd) $b$ quark
jet is expected to be hard for the signal (\ref{eq22},\ref{eq26}), but
soft for the background processes (\ref{eq27}). A $\ p_T > 80$ GeV cut
on this $b$-jet improves the signal/background ratio. Finally this
$b$-jet is combined with each of the reconstructed top pair to give
two entries of $M_{tb}$ per event. For the signal events, one of them
corresponds to the $H^\pm$ mass while the other constitutes a
combinatorial background. Figure 5 shows this invariant mass
distribution for the signal along with the above mentioned background
processes for different $H^\pm$ masses at $\tan \b = 40$ Similar
results hold for $\tan \b \simeq 1.5$. One can check that the
significance level of the signal is $S/\sqrt{B} \gsim 5$ \cite{twentyfive}. The
corresponding $H^\pm$ discovery reaches in the high and low $\tan \b$
regions are shown in Fig. 2. While the discovery reach via $tb$ is
weaker than that via the $\t\n$ channel in the high $\tan \b$ region,
the former offers the best $H^\pm$ discovery reach in the low $\tan
\b$ region. This is particularly important in view of the fact that
the indirect LEP limit shown in Fig. 2 gets significantly weaker with
the reported increase in the top quark mass, as discussed
earlier. 
 
One can also use 4 $b$-tags to look for the $H^\pm \to tb$ signal
\cite{twentysix}. The signal comes from (\ref{eq23},\ref{eq26}), and the
background from the first process of (\ref{eq27}). After the
reconstruction of the $t\bar t$ pair, both the remaining pair of
$b$-jets are expected to be soft for the background, since they come
from gluon splitting. For the signal, however, one of them comes from
the $H^\pm$ decay (\ref{eq26}); and hence expected to be hard and
uncorrelated with the other $b$-jet. Thus requiring a $p_T >$ 120 GeV
cut on the harder of the two $b$-jets along with large invariant mass
$(M_{bb} > 120$ GeV) and opening angle (cos$\theta_{bb} < 0.75$) for the
pair, one can enhance the signal/background ratio
substantially. Unfortunately the requirement of 4 $b$-tags makes the
signal size very small. Moreover the signal contains one soft $b$-jet
from (\ref{eq23}), for which one has to reduce the $p_T$ threshold from
30 to 20 GeV. The resulting signal and background cross-sections are
shown in Fig. 6 for $\tan \b = 40$. In comparison with Fig. 5 one can
see a significant enhancement in the signal/background ratio, but at
the cost of a much smaller signal size. Nonetheless this can be used
as a supplementary channel for $H^\pm$ search, provided one can
achieve good $b$-tagging for $p_T \sim 20$ GeV jets.

\subsection{Heavy $H^\pm$ Search in the $Wh^0$ Channel}

The LEP
limit of $M_{h^0} \gsim 100 $ GeV in the MSSM implies that the $H^\pm
\to Wh^0$ decay channel has at least as high a threshold as the $tb$
channel. The maximum value of its decay BR,
\be
B^{\rm max} (H^\pm \to Wh^0) \simeq 5 \% ,
\label{eq29}
\ee
is reached for $H^\pm$ mass near this threshold and low $\tan\b$. The
small BR for this decay channel is due the suppression of the
$H^+W^-h^0$ coupling relative to the $H^+ \bar t b$ coupling
(\ref{eq10}). Note that both the decay channels correspond the same
final state, $H^\pm \to b\bar b W$, along with an accompanying top
from the production process (\ref{eq22}). Nonetheless one can
distinguish the $H^\pm \to Wh^0$ from the $H^\pm \to tb$ as well as
the corresponding backgrounds (\ref{eq27}) by looking for a clustering
of the $b\bar b$ invariant mass around $M_{h^0}$ along with a veto on
the second top \cite{twentyseven}. Unfortunately the BR of
(\ref{eq29}) is too small to give a viable signal for this decay
channel. Note however that the LEP limit of $M_{h^0} \gsim 100$ GeV does
not hold in the CP violating MSSM \cite{five} or the singlet
extensions of the MSSM Higgs sector like the NMSSM
\cite{six}. Therefore it is possible to have a $Wh^0$ threshold
significantly below $m_t$ in these model. Consequently one can have a
$H^\pm$ boson lighter than the top quark in these models in the low
$\tan \b$ region, which can dominantly decay into the $Wh^0$
channel. Thus it is possible to have spectacular $t \to bH^+ \to
bWh^0$ decay signals at LHC in the NMSSM \cite{twentyseven} as well as
the CP violating MSSM \cite{twentyeight}.

\section{Concluding Remarks}

Let me conclude by commenting on a few aspects of $H^\pm$ boson
search, which could not be discussed in this brief review. The
associated production of $H^\pm$ with $W$ boson has been investigated
in \cite{twentynine}, and the $H^\pm H^\mp$ and $H^\pm A^0$
productions in \cite{thirty}. Being second order electroweak
processes, however, they give much smaller signals than (\ref{eq22}),
while suffering from the same background. However one can get
potentially large $H^\pm$ signal from the decay of strongly produced
squarks and gluinos at LHC, which can help to fill in the gap in the
intermediate $\tan\b$ region of Fig. 2 for favourable SUSY parameters
\cite{thirtyone}.

Finally, the SUSY quantum correction to 
$H^\pm$ production can be potentially important since it is known to
be nondecoupling, i.e. it remains finite even for very large SUSY mass
parameters \cite{thirtytwo, thirtythree, thirtyfour}. A brief
discussion of this effect can be found in a larger version of this
review \cite{thirtyfive}, which also covers $H^\pm$ search at LEP and
Tevatron.

\newpage

\begin{figure*}
\centerline{\epsfig{file=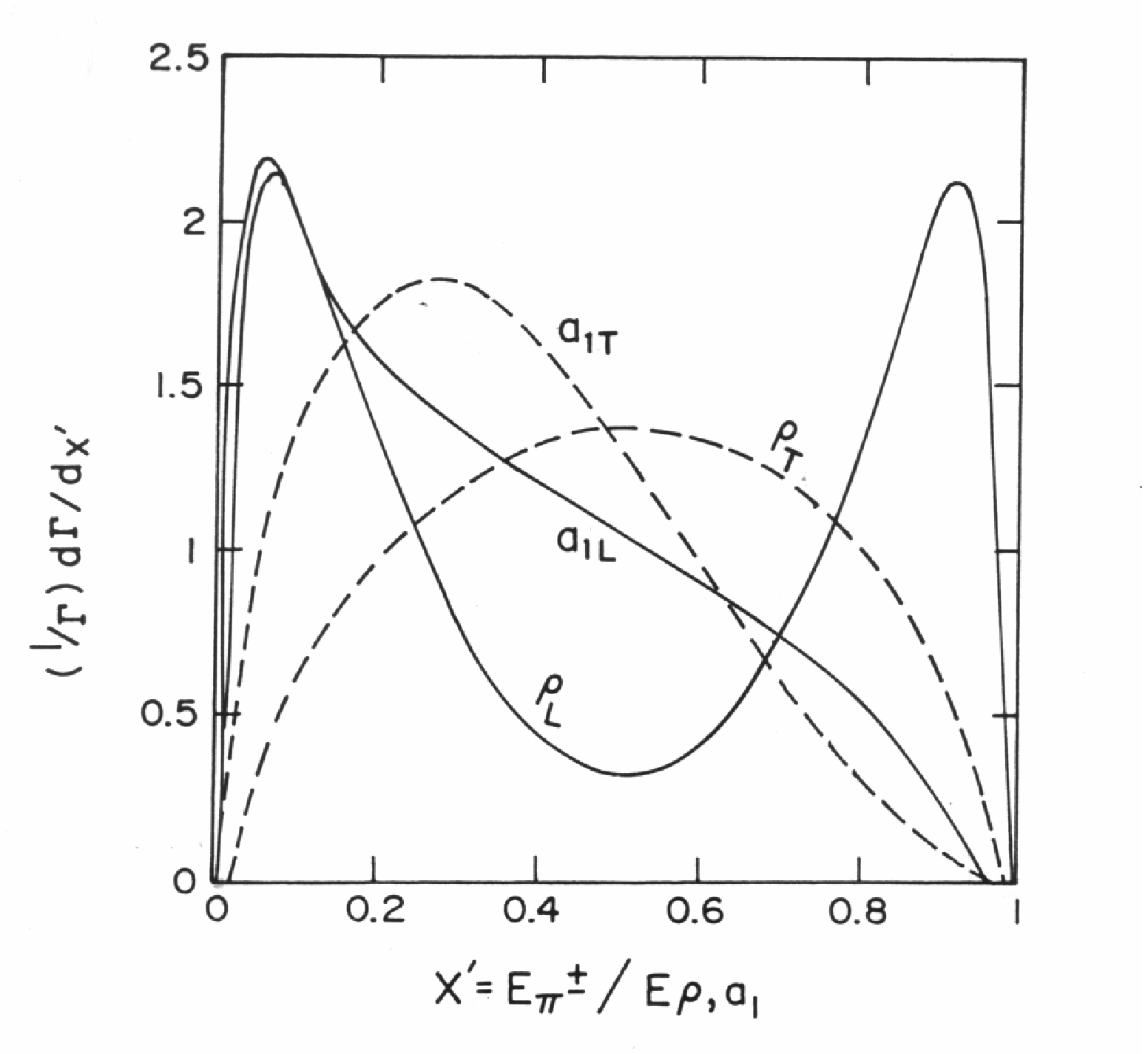,height=9cm,width=9cm,angle=0}}
\caption{\label{fig3} 
Distributions of the normalised decay widths of $\t^\pm$ via
$\r^\pm_{L,T} \to \p^\pm \p^0$ and $a^\pm_{1L,T} \to \p^\pm \p^0 \p^0$
in the momentum fraction carried by the charged pion
\cite{seventeen}. On this plot the $\t^\pm \to \p^\pm \n$ decay would
correspond to a $\delta$-function at $x' = 1$. }
\end{figure*}

\begin{figure*}
\centerline{\epsfig{file=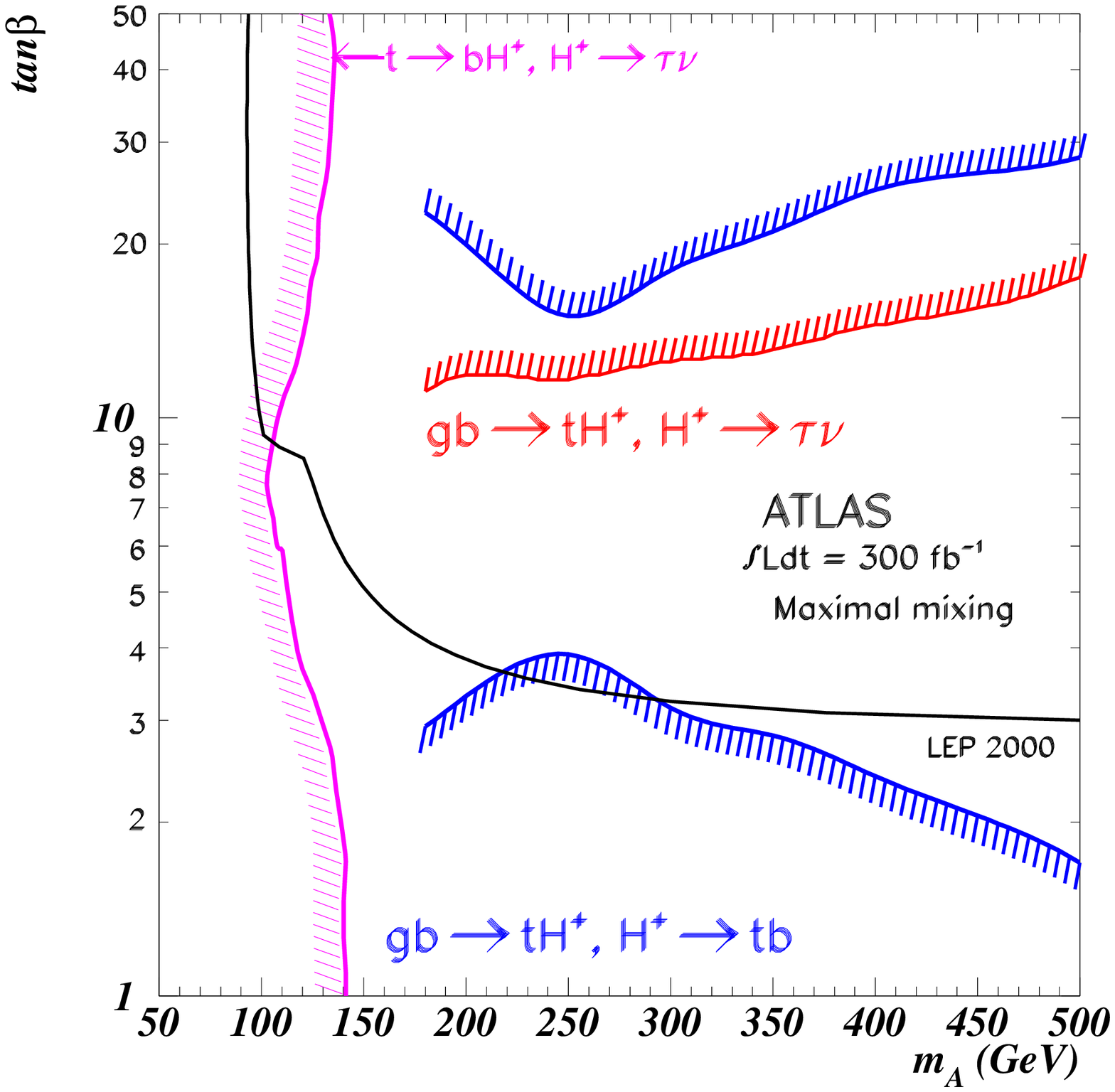,height=8cm,width=8cm,angle=0}}
\caption{\label{fig4} 
The 5-$\s$ $H^\pm$ boson discovery contours of the ATLAS experiment at
LHS from $t \to bH^+, H^+ \to \t\n$ (vertical); $gb \to tH^-, H^-
\t\n$ (middle horizontal) and $gb \to tH^-, H^- \to \bar tb$ (upper
and lower horizontal) channels \cite{nineteen}. One can see similar
contours for the CMS experiment in the second paper of
ref.\cite{nineteen}. The horizontal part of indirect LEP limit shown
here has weakened significantly now as explained in the text. }
\end{figure*}

\begin{figure*}[ht]
\begin{center}
\leavevmode
\epsfig{file=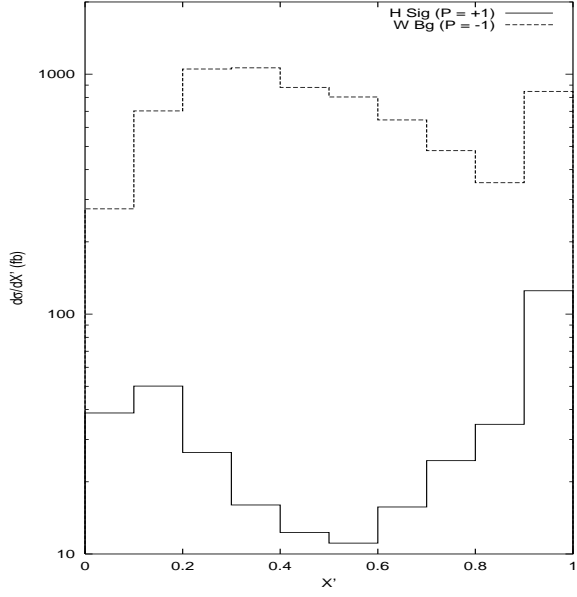,height=8cm,width=8cm,angle=0}
\end{center}
\caption{\label{fig5} 
The LHC cross-section for a 300 GeV $H^\pm$ signal at $\tan\b = 40$ 
shown along with the $t\bar t$ background in the 1-prong $\t$-jet channel,
as functions of the $\t$-jet momentum fraction carried by the charged pion.}
\end{figure*}

\begin{figure*}[hb]
\begin{center}
\leavevmode
\epsfig{file=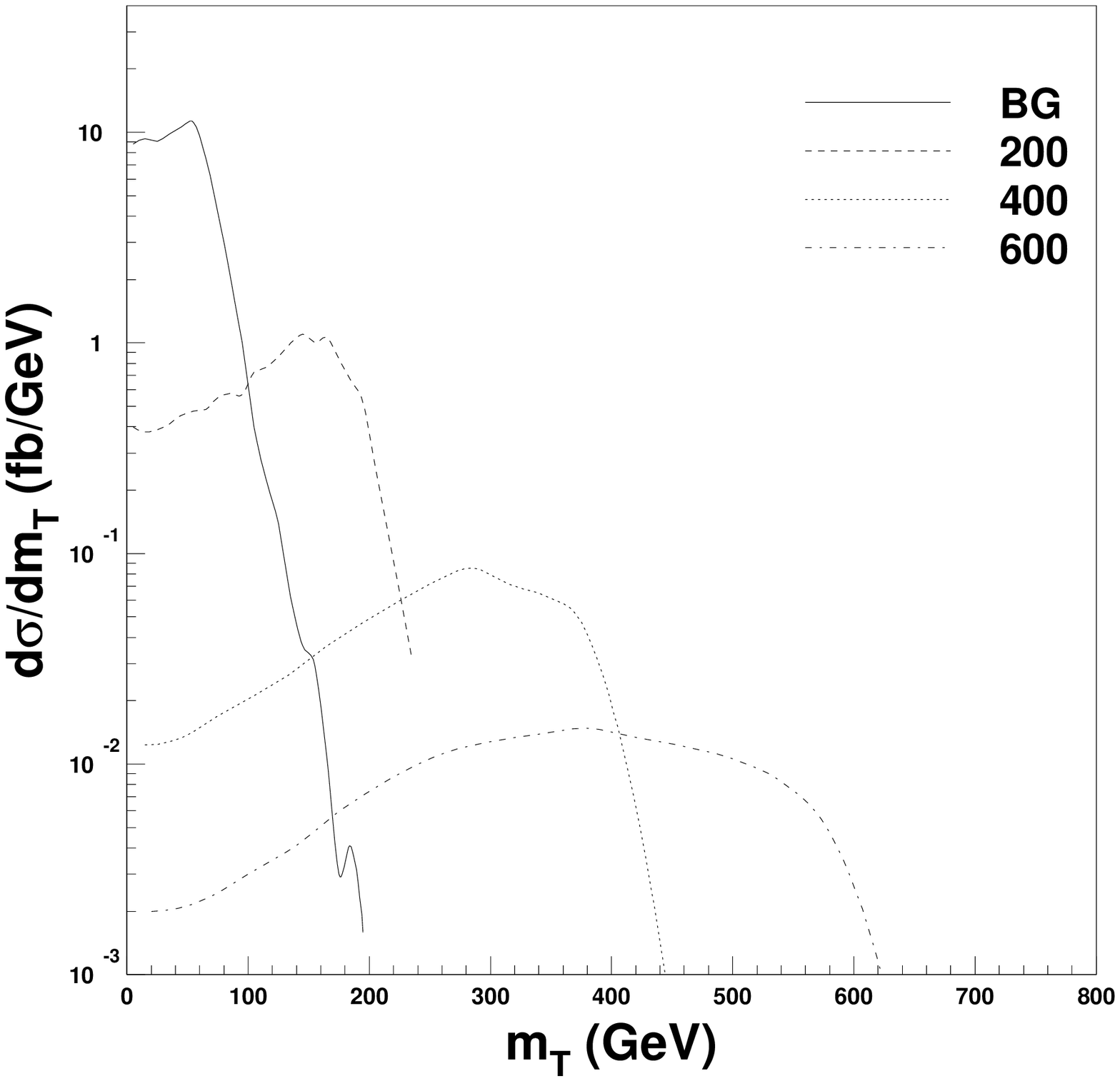,height=8cm,width=8cm,angle=0}
\epsfig{file=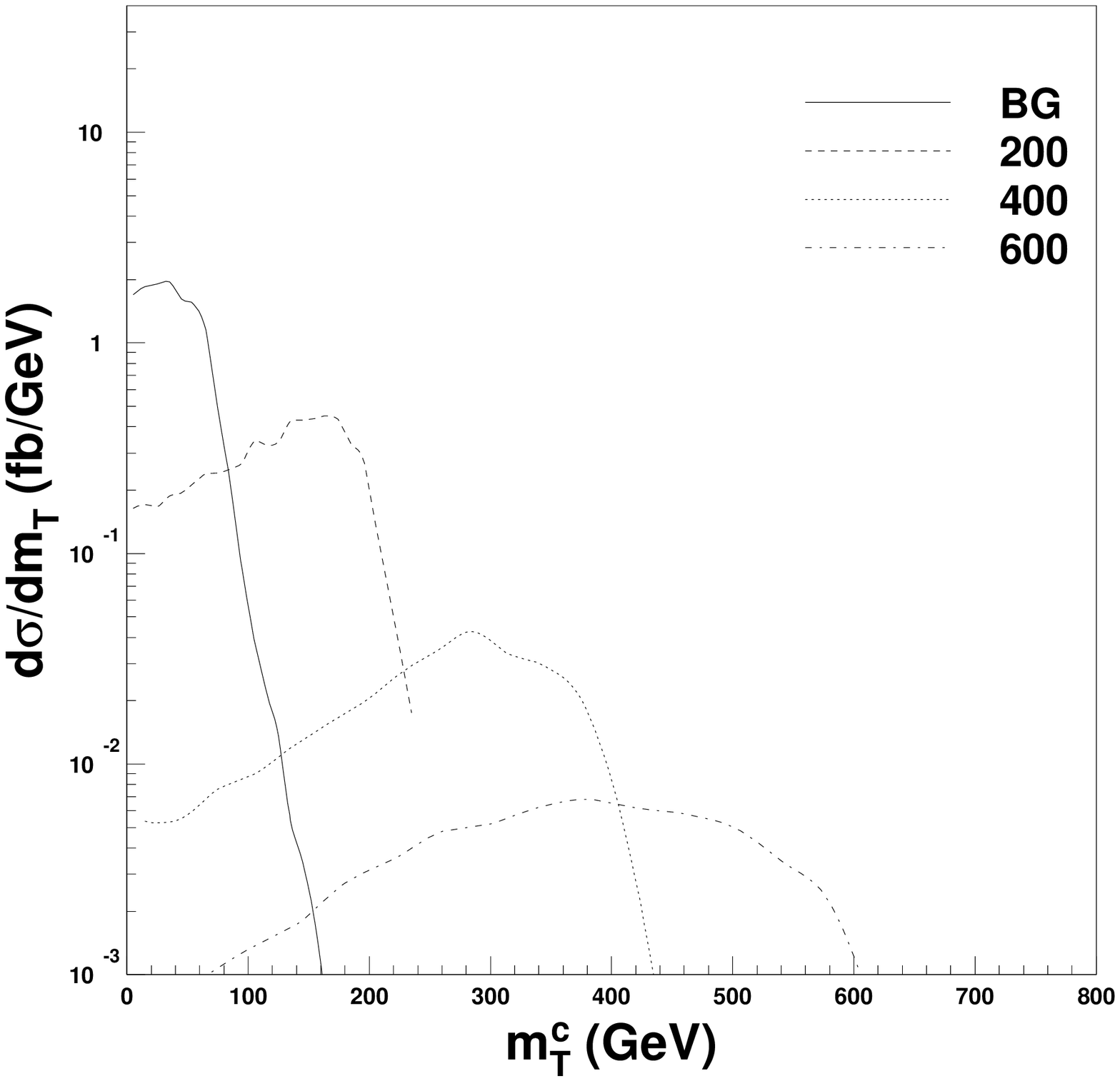,height=8cm,width=8cm,angle=0}
\end{center}
\caption{\label{fig6} 
Distributions of the $H^+$ signal and the $t\bar t$ background
cross-sections in the transverse mass of the $\t$-jet with
$p\!\!\!/_T$ for (left) all 1-prong $\t$-jets, and (right) those with
the charged pion carrying $> 80\%$ of the $\t$-jet momentum
($M_{H^\pm}$ = 200,400,600 GeV and $\tan\b$ = 40) \cite{twentythree}.}
\end{figure*}

\begin{figure*}[ht]
\begin{center}
\leavevmode
\epsfig{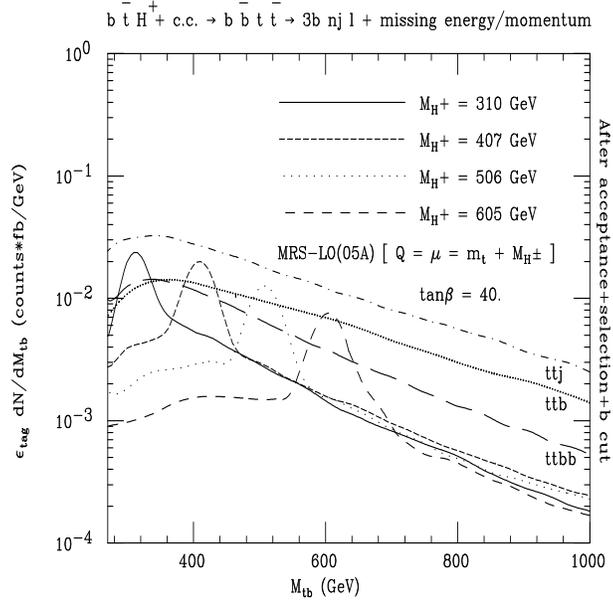}
\end{center}
\caption{\label{fig7} 
The reconstructed $tb$ invariant mass distribution of the $H^\pm$
signal and different QCD backgrounds in the isolated lepton plus
multijet channel with 3 $b$-tags \cite{twentyfive}.}
\end{figure*}

\begin{figure*}[hb]
\begin{center}
\leavevmode
\epsfig{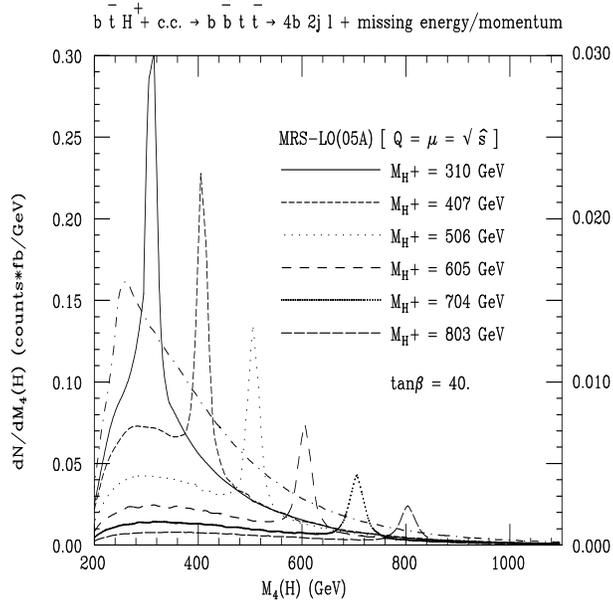}
\end{center}
\caption{\label{fig8} 
The reconstructed $tb$ invariant mass distribution of the $H^\pm$
signal and the QCD background in the isolated lepton plus multijet
channel with 4 $b$-tags \cite{twentysix}. The scale on the right
corresponds to applying a $b$-tagging efficiency factor $\e^4_b =
0.1$.}
\end{figure*}

\end{document}